\documentclass[traditabstract]{aa}
\usepackage{graphicx}
\usepackage{txfonts}
\usepackage[authoryear]{natbib}
\begin{document}

\title{Gamma Rays from Annihilations at the Galactic Center \\ in a Physical Dark Matter Distribution}

\author{A. Lapi\inst{1,2}, A. Paggi\inst{1}, A. Cavaliere\inst{1}, A.
Lionetto\inst{1,3}, A. Morselli\inst{1,3}, \and V.
Vitale\inst{1}} \institute{$^1$ Dip. Fisica, Univ. `Tor
Vergata', Via Ricerca Scientifica 1, I-00133 Roma, Italy.\\
$^2$ SISSA/ISAS, Via Beirut 2-4, I-34151 Trieste, Italy. \\
$^3$ INFN-Sezione di Roma2, Via Ricerca Scientifica 1, I-00133
Roma, Italy.}

\date{\today}

\abstract{We discuss the $\gamma$-ray signal to be expected
from dark matter (DM) annihilations at the Galactic Center. To
describe the DM distribution in the Galactic halo we base on
the Jeans equation for self-gravitating, anisotropic
equilibria. In solving the Jeans equation, we adopt the
specific correlation between the density $\rho(r)$ and the
velocity dispersion $\sigma^2_r(r)$ expressed by the powerlaw
behavior of the DM `entropy' $K\equiv
\sigma_r^2/\rho^{2/3}\propto r^\alpha$ with $\alpha\approx
1.25-1.3$. Indicated (among others) by several recent $N$-body
simulations, this correlation is privileged by the form of the
radial pressure term in the Jeans equation, and yields a main
body profile consistent with the classic self-similar
development of DM halos. In addition, we require the Jeans
solutions to satisfy regular boundary conditions both at the
center (finite pressure, round gravitational potential) and in
the outskirts (finite overall mass). With these building blocks
we derive physical solutions, dubbed `$\alpha$-profiles'. We
find the one with $\alpha=1.25$, suitable for the Galaxy halo,
to be intrinsically flatter at the center relative to the
empirical NFW formula, yet steeper than the empirical Einasto
profile. So on scales of $10^{-1}$ deg it yields annihilation
fluxes \emph{lower} by a factor $5$ than the former yet
\emph{higher} by a factor $10$ than the latter; such fluxes
will eventually fall within the reach of the \textsl{Fermi}
satellite. We show the effectiveness of the $\alpha$-profile in
relieving the astrophysical uncertainties related to the
macroscopic DM distribution, and discuss its expected
performance as a tool instrumental to interpret the upcoming
$\gamma$-ray data in terms of DM annihilation.}


\authorrunning{A. Lapi et al.}
\titlerunning{$\gamma$ rays from DM annihilations at the GC}

\maketitle

\section{Introduction}

Several astrophysical and cosmological probes \citep[for a
review see][]{Bertone2005} have firmly established that baryons
-- which stars, planets, and (known) living creatures are made
of -- constitute only some $15\%$ of the total \emph{matter}
content in the Universe adding to the dominant dark energy
component. The rest is in the form of `cold dark matter' (DM),
i.e., massive particles that were non-relativistic at
decoupling, do not emit/absorb radiations, and basically do not
interact with themselves and with the baryons except via
long-range gravitational forces.

However, no `direct' detection of the DM has been made so far,
other than \citet{Bernabei2009}. Thus the microscopic nature of
the DM remains largely a mystery; several clues suggest as a
promising candidate or component the lightest supersymmetric
particle, the `neutralino' \citep[for a review
see][]{Bertone2009book}. Given that the latter's mass,
depending on the specific supersymmetric model, ranges from
several GeVs to tens of TeVs, its laboratory production
requires an accelerator at least as powerful as the newly-born
Large Hadron Collider \citep[see][]{Baer2009}; the discovery of
supersymmetry and specifically of the neutralino is one of the
main aims for the current experiments in high-energy physics.

Meanwhile, evidence for the DM can be looked for `indirectly'
in the sky. In fact, the basic aims of the recently launched
\textsl{Fermi} satellite include the search for $\gamma$-ray
signals due to the annihilation of DM particles at the Galactic
Center (GC) and in nearby galaxies (see discussion in \S~4).
The former provides a favorable target being closest to us,
with the DM density expected to increase in moving toward the
inner regions of a galaxy. However, the GC is also a crowded
region, and it remains a challenging task to separate the DM
signal from the contributions of other astrophysical sources
and backgrounds whose energy spectrum and angular distribution
are poorly known.

In principle, if one can predict the strength and angular
distribution of the annihilation signal itself, then the
$\gamma$-ray observations would elicit, or put `indirect'
constraints on the (combined) \emph{microscopic} properties of
the DM particles like mass, annihilation cross section and
channels. This approach has been pursued extensively
\citep[e.g.,][]{Bergstrom1998, Fornengo2004, Strigari2007,
Bertone2009, Serpico2009} but suffers yet of large
uncertainties \citep[see][]{Cesarini2004}, mainly related to
the poor knowledge of the macroscopic DM distribution $\rho(r)$
throughout the Galaxy.

Since the annihilation rate scales like $\rho^2(r)$, such
uncertainties are maximized near the center right where
detection is favored. Note that similar if milder uncertainties
affect the source function of the electrons originated from DM
annihilations by production or cascading; these diffuse
outwards and interact with the Galactic magnetic field and with
the interstellar light to produce synchrotron emission observed
in the radio band \cite[see][]{Bertone2009}, and inverse
Compton radiation observable in $\gamma$ rays
\cite[see][]{Papucci2009}.

Traditionally, the density profile of an equilibrium DM
structure, or `halo', is rendered in terms of different
empirical formulas that fit the results of $N$-body simulations
and to some extent the stellar observations. Perhaps the most
popular one is the Navarro, Frenk \& White \citep[hereafter
NFW; see][]{Navarro1997} profile, that has an asymptotic inner
slope $\rho(r)\propto r^{-1}$, goes over to a powerlaw behavior
$\rho(r)\propto r^{-2}$ in the halo's middle, and declines as
$\rho(r)\propto r^{-3}$ in the outer regions. Despite its
widespread use in the literature, clearly this expression
cannot account for the actual DM distribution in the inner
regions of a galaxy halo where it would imply a centrally
angled gravitational potential well and an infinite pressure,
nor in the halo outskirts where it would yield a diverging
overall mass.

Other empirical density profiles have been proposed but suffer
of similarly unphysical features; e.g., the Moore profile
\citep[see][]{Diemand2005} goes like $\rho(r)\propto r^{-1.2}$
and implies a gravitational force diverging towards the center,
while the Einasto profile \citep[see][]{Graham2006} behaves
like $\rho(r)\propto e^{-a\,r^{b}}$, so it yields a vanishing
pressure there. We stress that the \emph{differences} in the
predicted annihilation signals under these DM distributions
turn out to be quite considerable; for example, the ratio of
the NFW to the Einasto squared density averaged over $1$ degree
(about $150$ pc) comes to a factor $10$ when normalized at the
Sun's location (see also discussion in \S~4).

Our stand here is that the macroscopic uncertainties yielding
such differences can, and ought to be relieved. To this
purpose, in \S~2 we present the \emph{physical} density
distributions that we dub \emph{$\alpha$-profiles}; these are
solutions of the Jeans equation that satisfy regular inner and
outer boundary conditions. In \S~3 we use the $\alpha$-profile
suitable for the Galaxy halo as the macroscopic benchmark to
evaluate the DM annihilation signal expected from the GC. As
for the microscopic sector, we base on a standard model for the
mass, cross section and annihilation channel of the DM
particles, the extension to more complex microphysics being
straightforward. Finally, our findings are summarized and
discussed in \S~4.

Throughout this work we adopt a standard, flat cosmology
\cite[see][]{Dunkley2009} with normalized matter density
$\Omega_M = 0.27$, and Hubble constant $H_0 = 72$ km s$^{-1}$
Mpc$^{-1}$.

\section{Development and structure of DM halos}

Galaxies are widely held to form under the drive of the
gravitational instability that acts on initial perturbations
modulating the cosmic density of the dominant cold DM
component. At first the instability is kept in check by the
cosmic expansion, but when the local gravity prevails collapse
sets in, and form a DM halo in equilibrium under self-gravity.
The amplitude of more massive perturbations is smaller, so the
formation is progressive in time and hierarchical in mass, with
the largest structures forming typically later \citep[see][for
a review]{Peebles1993}.

\subsection{Two-stage evolution}

Such a formation history has been resolved to a considerable
detail by many $N$-body simulations \citep[e.g.,][]{White1986,
Springel2006}; recently, a novel viewpoint emerged.

Firstly, the halo growth has been recognized
\citep[see][]{Zhao2003, Wechsler2006, Hoffman2007, Diemand2007}
to comprise two stages: an early fast collapse including a few
violent major mergers, that builds up the halo main `body' with
structure set by dynamical relaxation; and a later,
quasi-equilibrium stage when the body is nearly unaffected,
while the outskirts develop from the inside-out by minor
mergers and smooth accretion \citep[see][]{Salvador2007}. The
\emph{transition} is provided by the time when a DM
gravitational well attains its maximal depth, i.e., the radial
peak of the circular velocity $v^2_c\equiv G\, M/R$ attains its
maximal height, along a given growth history
\citep[see][]{Li2007}.

Secondly, generic features of the ensuing equilibrium
structures have been sought \citep[see][]{Hansen2004,
Dehnen2005, Schmidt2008} among powerlaw correlations of the
form $\sigma_D^{2\epsilon/3}/\rho^{2/3}\propto r^\alpha$; this
involves the density $\rho(r)$ and the velocity dispersion
$\sigma_D^{2}\equiv \sigma_r^{2}\,(1+D\,\beta)$, with
anisotropy inserted via the standard \citet{Binney1978}
parameter $\beta\equiv 1-\sigma_\theta^2/\sigma_r^2$ and
modulated by the index $D$ \citep[see][]{Hansen2007}. It is
matter of debate which of these correlations best apply, see
\citet{Schmidt2008} and \citet{Navarro2008}; the former
authors, in particular, find that the structure of different
simulated halos may be described by different values of $D$,
with linearly related values of $\epsilon$ and $\alpha$ (see
their Eqs.~4 and 5).

Here we shall focus on the specific correlation
\begin{equation}
K\equiv {\sigma_r^2\over \rho^{2/3}}\propto r^\alpha~
\end{equation}
that involves solely the squared radial dispersion
$\sigma_r^2$, corresponding to $D=0$ and $\epsilon=3$. This is
because $K$ has not only the striking form of a DM `entropy'
(or rather adiabat), but also the related operational advantage
of providing a \emph{direct} expression of the radial pressure
term $\rho\,\sigma_r^2=K\,\rho^{5/3}\propto
r^\alpha\,\rho^{5/3}$ appearing in the Jeans equation for the
radial equilibrium; in the latter any anisotropy is already
accounted for by a separate term (see Eq.~2 below). On the
other hand, the correlation $K\propto r^\alpha$ with
$\alpha\approx 1.25-1.3$ provides a simple yet effective fit of
many simulations \citep[see][and many others]{Taylor2001,
Rasia2004, Hoffman2007, Diemand2007, Schmidt2008,
Ascasibar2008, Navarro2008, Vass2008}. In the lower $\alpha$
range, Eq.~(1) has the added bonus of preserving the classic
self-similar slope in the halo body (see Eq.~3 below).

To independently probe the matter, \citet{Lapi2009a} performed
a semianalytical study of the two-stage halo development, and
derived (consistently with the simulations) that $\alpha$ is
set at the transition time via scale-free stratification of the
particle orbits throughout the halo body, and thereafter
remains closely constant and uniform at a value within the
narrow range $1.25 - 1.3$. Moreover, they found that \emph{on
average} the values of $\alpha$ depend though weakly on the
mass of the halo, such that $\alpha\approx 1.3$ applies to
galaxy clusters, while $\alpha\approx 1.25$ applies to Milky
Way sized galaxies.

\subsection{The DM $\alpha$-profiles}

The halo physical profiles may be derived from the radial Jeans
equation, with the radial pressure $\rho\sigma^2_r \propto
r^{\alpha}\, \rho^{5/3}$ and anisotropies described by the
standard \citet{Binney1978} parameter $\beta$. Thus the Jeans
equation simply writes
\begin{equation}
\gamma = {3\over 5}\,\left(\alpha+{v_c^2\over \sigma_r^2}\right) + {6\over
5}\,\beta~
\end{equation}
in terms of the logarithmic density slope $\gamma\equiv
-\mathrm{d}\log\rho/\mathrm{d}\log r$. As first shown by
\citet{Austin2005} and \citet{Dehnen2005}, Jeans supplemented
with the mass definition $M(<r)\equiv
4\pi\int_0^r{\mathrm{d}r'}~r'^2\,\rho(r')$ entering
$v_c^2\equiv G M(<r)/r$, provides an integro-differential
equation for $\rho(r)$, that by double differentiation reduces
to a handy $2^{\mathrm{nd}}$-order differential equation for
$\gamma$.

To set the context for the Milky Way DM distribution, we recall
that the space of solutions for Eq.~(2) spans the range
$\alpha\leq 1.\overline{296}$; the one for the upper bound and
the behaviors of others ones have been analytically
investigated by \citet{Austin2005} and \citet{Dehnen2005}. In
\citet{Lapi2009a} we explicitly derive the Jeans solutions with
$\beta=0$ (meaning isotropy) for the full range $\alpha\approx
1.25-1.\overline{296}$ subjected to regular boundary conditions
both at the center and in the outskirts, i.e., a \emph{round}
minimum of the potential with a \emph{finite} pressure (or
energy density), and a \emph{finite} (hence definite) overall
mass, respectively. These we dubbed `$\alpha$-\emph{profiles}'.

The corresponding density runs steepen \emph{monotonically}
outwards, and are summarized by the pivotal slopes
\begin{equation}
\gamma_a \equiv {3\over 5}\,\alpha~,~~~~~~~~ \gamma_0\equiv
6-3\alpha~,~~~~~~~~ \gamma_b\equiv {3\over2}\,(1+\alpha)~;
\end{equation}
these start from the central ($r\rightarrow 0$) value
$\gamma_a\approx 0.75-078$, steepen in the halo main body to
$\gamma_0\approx 2.25-2.1$ (the former being the slope from the
classic self-similar collapse), and steepen further into the
outskirts to typical values $\gamma_b\approx 3.38-3.44$ before
a final cutoff. Thus the inner slope is considerably
\emph{flatter} and the outer slope \emph{steeper} compared to
the empirical NFW formula \citep[see][]{Navarro1997}; in
comparison to the Einasto profile, the main difference occurs
in the inner regions where the $\alpha$-profile rather than
flat is moderately steep (see Eq.~3 and Fig.~1).

For a density profile, a relevant parameter is the
`concentration' $c\equiv R_v/r_{-2}$, defined in terms of the
virial radius $R_v$ and of the radius $r_{-2}$ where
$\gamma=2$; in the context of $\alpha$-profiles $c$ may be
viewed as a measure either of central condensation (small
$r_{-2}$) or of outskirts' extension (large $R_v$). The
concentration constitutes an indicator of the halo age; in
fact, numerical experiments \citep[see][]{Bullock2001,
Zhao2003, Wechsler2006, Diemand2007} show that $c(z)\approx
3.5$ holds at the end of the fast collapse stage, to grow as
$c(z)\approx 3.5\,(1+z_t)/(1+z)$ during the slow accretion
stage after the transition at $z_t$. Current values $c\approx
3.5\,(1+z_t)\approx 10$ apply for a galaxy like the Milky Way
that had its transition at $z_t\lesssim 2$.

\begin{figure}
\includegraphics[width=10cm]{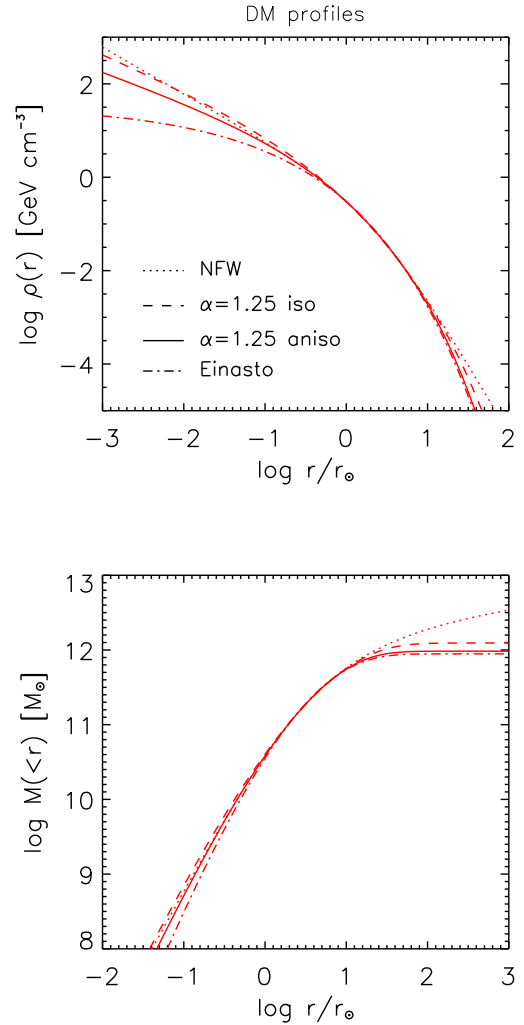}
\caption{Density and mass profiles in the Milky Way. The dashed
and solid lines illustrate the $\alpha$-profiles with
$\alpha=1.25$ in the isotropic and the anisotropic case, with
$\gamma_a=0.75$ and $0.63$, respectively; the dotted line
represents the NFW formula, and the dot-dashed line refers to
the Einasto profile. All profiles are normalized to the local
density $0.3$ GeV cm$^{-3}$ at the Sun's location
$r_{\odot}\approx 8.5$ kpc within the Galaxy; we have adopted
$r_{-2} = 20$ kpc and $c = 10$, see \S~2.}
\end{figure}

The density and mass distribution in the Milky Way are
illustrated in Fig.~1 for the isotropic $\alpha$-profiles with
$\alpha=1.25$ (dashed), for the NFW formula (dotted), and for
the Einasto profile (dot-dashed). All densities have been
normalized to the local density $0.3$ GeV cm$^{-3}$ at the
Sun's location $r_{\odot}\approx 8.5$ kpc within the Galaxy. We
further adopt $r_{-2} = 20$ kpc (consistent with $c = 10$).
Note from Fig.~1 that the Einasto and NFW profiles
\emph{differ} substantially at the center as for the density,
and in the outskirts as for the mass, while the
$\alpha$-profile strikes an \emph{intermediate} course between
the two.

\subsection{Anisotropy}

It is clear from Eq.~(2) that anisotropy will \emph{steepen}
the density run for positive $\beta$ meaning radial velocity
dominance, as expected in the \emph{outskirts} from infalling
cold matter. On the other hand, tangential components
(corresponding to $\beta\lesssim 0$) must develop toward the
center, as expected from increasing importance of angular
momentum effects. This view is supported by numerical
simulations \citep[see][]{Austin2005, Hansen2006, Dehnen2005},
which in detail suggest the effective linear approximation
\begin{equation}
\beta(r)\approx \beta(0)+\beta'\,[\gamma(r)-\gamma_a]
\end{equation}
to hold with $\beta(0)\geq -0.1$ and $\beta'\approx 0.2$,
limited to $\beta(r)< 0.5$.

In \citet{Lapi2009b} we extended the $\alpha$-profiles to such
anisotropic conditions in the full range $\alpha\approx
1.25-1.3$, inspired by the analysis by \cite{Dehnen2005} for
the specific case $\alpha\approx 1.3$. We find the
corresponding $\rho(r)$ to be somewhat \emph{flattened} at the
center by a weakly negative $\beta(0)$, and further
\emph{steepened} into the outskirts where $\beta(r)$ grows
substantially positive. Specifically, the following simple
rules turn out to apply: the slope $\beta'$ in Eq.~(4) drops
out from the derivatives of the Jeans equation
\citep[see][]{Dehnen2005}; the upper bound to $\alpha$ now
reads $35/27-4\beta(0)/27$; moreover, $\gamma_a$ is modified
into $3\alpha/5+6\beta(0)/5$ while $\gamma_0$ and $\gamma_b$
retain their form.

The anisotropic $\alpha$-profiles for the Milky Way are shown
as solid lines in Fig.~1. We note, in particular, that even a
limited central anisotropy (e.g., $\beta[0]\approx -0.1$)
causes an appreciable \emph{flattening} of the inner density
slope bringing it down to $\gamma_a\approx 0.63$ for
$\alpha=1.25$. This, of course, results in an even more
considerable flattening for the slope of the squared density,
the relevant quantity in our context of DM annihilations.

\subsection{A guide to profile computations}

Finally, in the Appendix we provide user-friendly analytic fits
for the density runs of the $\alpha$-profiles in terms of
standard deprojected S\'ersic formulas, but with parameters
directly \emph{derived} from the Jeans equation.

We stress that these physical $\alpha$-profiles with their
analytic fits are relevant to, and recently tested in several
contexts, including the interpretation of gravitational lensing
observations \citep[see][]{Lapi2009b}, the physics of the hot
diffuse baryons constituting the Intra-Cluster Plasma
\citep[see][]{Cavaliere2009}, and galaxy kinematics
\citep[see][]{Lapi2009c}. In the following we focus on the
specific $\alpha$-profile with $\alpha=1.25$ suitable for the
Milky Way halo (see \S~2.1) to predict the DM annihilation
signal from the GC.

\section{$\gamma$-ray signal from DM annihilation at the GC}

The $\gamma$-ray flux per solid angle due to DM annihilation
along a direction at an angle $\psi$ relative to the l.o.s.
toward the GC may be written (under the commonly assumed
spherical symmetry) as
\begin{equation}
{{\rm d} \Phi_\gamma\over {\rm d}\Omega}=3.74\times
10^{-6}\,N_{\gamma}\,\left({\langle\Sigma v\rangle\over
10^{-26}\,\mathrm{cm}^3\,\mathrm{s}^{-1}}\right)\,\\
\left({m_{\rm DM}\over 50\, \mathrm{GeV}}\right)^{-2}\,
J(\psi)
\end{equation}
in units of $\mathrm{m}^{-2}$ $\mathrm{s}^{-1}$
$\mathrm{sr}^{-1}~$. The above expression is naturally
factorized into a \emph{microscopic} and an
\emph{astrophysical} term \citep[e.g.,][and references
therein]{Bergstrom2009}. The former involves the mass of the DM
particle $m_{\rm DM}$, the number of photons $N_{\gamma}$
created per annihilation, and the angle-velocity averaged
annihilation rate $\langle \Sigma v\rangle$ in terms of the
particles' cross section $\Sigma$ and velocity $v$.

For the sake of definiteness we begin from considering a
neutralino DM particle with mass $m_{DM}\approx 50$ GeV,
annihilating through the $b\bar{b}$ channel (with $100\%$
branching ratio). We use the benchmark value for the
annihilation rate $\langle \Sigma v\rangle\approx 3\times
10^{-26}$ cm$^3$ s$^{-1}$, corresponding to a thermal relic
with a density close to the cosmological DM abundance
\begin{equation}
\Omega_{\rm DM}\,h^2\approx {3\times 10^{-27}\, \mathrm{cm}^3\,
\mathrm{s}^{-1}\over \langle\Sigma v\rangle}\approx 0.1
\end{equation}
as measured by \textsl{WMAP} \citep[see][]{Dunkley2009}. To
compute $N_\gamma=\int{\rm d}E~
\mathrm{d}N_\gamma/\mathrm{d}E_\gamma$ we adopt a photon
annihilation spectrum with shape
\begin{equation}
{\mathrm{d}N_\gamma\over \mathrm{d}x}=\eta\,x^a\,e^{b+cx+dx^2+ex^3}~,
\end{equation}
obtained from extrapolating the results by \citet{Fornengo2004}
down to energies $E_\gamma\approx 200$ MeV; here $x\equiv
E_\gamma/m_{\rm DM}$ is the energy normalized to the DM mass,
while $\eta=1$, $a=-1.5$, $b=0.579$, $c=-17.6080$, $d=23.862$,
$e=-25.181$ are fitting parameters for the adopted microscopic
DM model (see above).

The astrophysical term of Eq.~(5) is given by the integral of
the (squared) DM density projected along the l.o.s.
\begin{equation}
J(\psi)=\int{{\rm d}\ell\over
r_{\odot}}~{\rho^2(r)\over\rho^2(r_{\odot})}~,
\end{equation}
normalized to $\rho(r_{\odot})\approx 0.3$ GeV cm$^{-3}$, the
local density at the Sun's location $r_{\odot}\approx 8.5$ kpc.
A non-trivial angular dependence results from the peripheral
position of the Sun within the Milky Way halo, and involves
only the angle $\psi$ between the observed direction of the sky
and the GC; in terms of Galactic latitude $b$ and longitude $l$
with $\cos\psi = \cos{b}\,\cos{l}$, the radial variable can be
expressed as
$r=(r_\odot^2+\ell^2-2\,r_\odot\ell\cos{\psi})^{1/2}$ on using
the distance $\ell$ along the l.o.s. Finally, when observing a
region at an angular resolution $\Delta\Omega$, one has to
consider the average value of $J$, namely,
\begin{equation}
\bar{J}=\int_{\Delta\Omega}{\rm d}\Omega~J(\psi)~,
\end{equation}
with ${\rm d}\Omega=\cos{b}\,{\rm d}b\,{\rm d}l$.

\begin{figure}
\includegraphics[width=9cm]{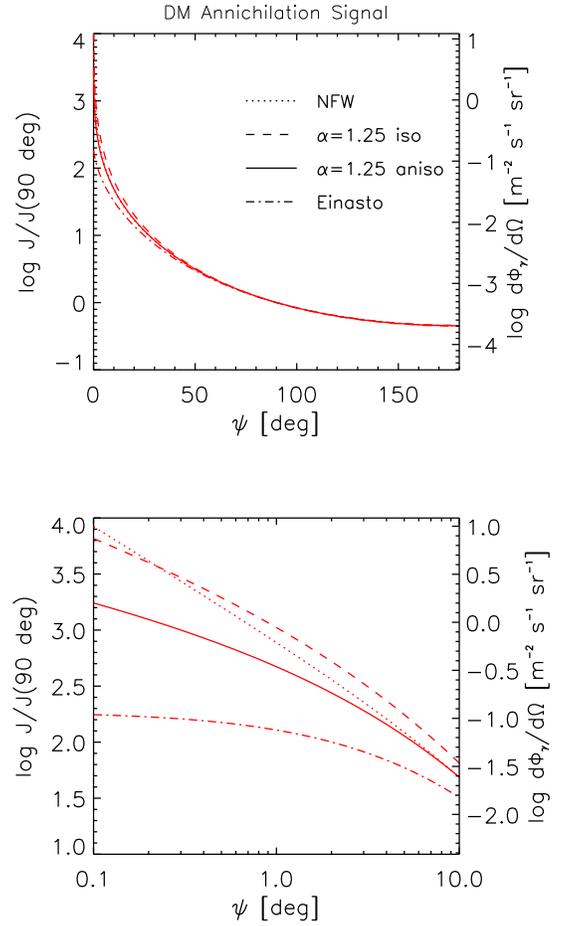}
\caption{The astrophysical factor $J(\psi)$ normalized to the
value at $\psi=90$ deg (left axis), and the corresponding
annihilation flux per unit solid angle for $E_{\gamma}\geq 200$
MeV (right axis; see \S~3 for details). The
bottom panel zooms on the inner angular scales.}
\end{figure}

We compute and report in Table~1 the values of $\bar{J}$ at
angular resolutions $\Delta\Omega = 10^{-3}$ sr and $10^{-5}$
sr for the $\alpha$-profile with $\alpha=1.25$ in the isotropic
and anisotropic cases, for the NFW formula, and for the Einasto
profile. In Table~2 we list the corresponding values of the
$\gamma$-ray flux for energies $E_\gamma\geq 200$ MeV. These
outcomes are illustrated in Fig.~2. It is seen that relative to
the NFW distribution, the fluxes predicted from the isotropic
and anisotropic $\alpha$-profile are lower by factors from a
few to several. Such fluxes are still within the reach of the
\textsl{Fermi} satellite; in fact, on the basis of the
simulations performed by \citet{Baltz2008},
\citet{Striani2009}, and \citet{Vitale2009}, we expect the
annihilation signal to be probed at a $3$-$\sigma$ confidence
level over a few years.

The above values may be compared with the current upper bound
to the integrated flux of $2.43^{+0.02}_{-0.02}\times 10^{-3}$
m$^{-2}$ s$^{-1}$ based on \textsl{Fermi} measurements at
$E_\gamma\geq 200$ MeV during $8$-month observations of the GC
over a solid angle $\Delta\Omega\approx 10^{-3}$ sr
\citep[see][]{Abdo2009, Atwood2009, Vitale2010}; this bound
decreases as $1/\sqrt{t}$ with the observation time $t$.
However, the flux currently observed includes contributions
from diffuse or not yet resolved Galactic sources, that are
being progressively removed \citep[see][]{Goodenough2009,
Striani2009, Vitale2010}; next stages of such a process will
take longer observations aimed at determining the spectrum of
individual resolved sources and a careful likelyhood analysis
of the backgrounds \citep[see discussion by][]{Cesarini2004}.

\begin{table}
\caption{Values of the astrophysical factor $\bar{J}$.}
\centering
\begin{tabular}{cccccc}
\hline\hline
DM profile & $\Delta\Omega=10^{-3}$  & $\Delta\Omega=10^{-5}$\\
\hline
\\
NFW & $1.2\times 10^3$ & $1.05\times 10^4$\\
\\
$\alpha$ iso & $1.3\times 10^3$ & $6.5\times 10^3$\\
\\
$\alpha$ aniso & $5.1\times 10^2$ & $1.6\times 10^3$\\
\\
Einasto & $1.1\times 10^2$ & $1.3\times 10^2$\\
\\
\hline
\end{tabular}
\end{table}

\begin{table}
\caption{Values of the $\gamma$-ray flux (in m$^{-2}$ s$^{-1}$)
for $E_\gamma\geq 200$ MeV.} \centering
\begin{tabular}{cccccc}
\hline\hline
DM profile & $\Delta\Omega=10^{-3}$ & $\Delta\Omega=10^{-5}$\\
\hline
\\
NFW & $4.7\times 10^{-4}$ & $4.0\times 10^{-5}$\\
\\
$\alpha$ iso & $5.2\times 10^{-4}$ & $2.5\times 10^{-5}$\\
\\
$\alpha$ aniso & $2.0\times 10^{-4}$ & $6.0\times 10^{-6}$\\
\\
Einasto & $4.1\times 10^{-5}$ & $5.0\times 10^{-7}$\\
\\
\hline
\end{tabular}
\end{table}

\section{Discussion and conclusions}

We have presented our $\alpha$-profile with $\alpha=1.25$ for
the equilibrium density and mass distributions in a galactic DM
halo, and specifically in the Milky Way. We have shown that
this profile constitutes the robust solution of the
\emph{equilibrium} Jeans equation with \emph{physical} inner
and outer boundary conditions, i.e., finite pressure and round
potential minimum at the center, and finite overall mass. The
corresponding density profile $\rho(r)$ is intrinsically
\emph{flatter} at the center, and intrinsically \emph{steeper}
in the outskirts, relative to the empirical NFW formula. These
features are \emph{sharpened} yet in halos with anisotropic
random velocities. We have also provided the reader with a
precise and user-friendly analytic fit to the $\alpha$-profile
(see Appendix for details).

Then we have focused on the role of this $\alpha$-profile as a
\emph{benchmark} for computing the DM annihilation signal
expected from the GC. In fact, we have computed the
`astrophysical factor' $J(\psi)$ (angular distribution,
independent of microphysics) entering the expression of the
annihilation flux. As a definite example, we have also computed
the $\gamma$-ray flux on adopting a simple, fiducial
microscopic model; this we find consistent with current
\textsl{Fermi} observations, given that the latter may include
contributions from still unresolved point sources.

Given the physical $\alpha$-profile and the corresponding
factor $J(\psi)$, the extension to more complex microscopic
scenarios like \textsl{mSUGRA} \citep[started
by][]{Chamseddine1982, Barbieri1982, Ohta1983, Hall1983} will
be easily made in terms of annihilations channels, cross
sections and particle masses. In this context our
$\alpha$-profile \emph{relieves} astrophysical uncertainties
related to the macroscopic DM distribution. We stress that
constraints on particle cross sections and masses inferred from
radio and $\gamma$-ray observations of the GC have been to now
more sensitive to the assumed DM distribution than to specific
annihilation channels (different from leptonic $\tau\bar\tau$),
see Figs.~3 and 4 in \citet{Bertone2009}. In fact, the latter
show that a DM distribution with an inner slope like our
$\alpha$-profile is required to allow cross sections
$\langle\Sigma\,v\rangle\ga 10^{-26}$ cm$^3$ s$^{-1}$ with
masses $m_{\rm DM}\la 500$ GeV for the non-leptonic channels
that are widely considered on grounds of theoretical
microphysics.

Concerning small scales $r\sim$ a few tens of pcs around the
GC, we touch upon a number of possible deviations of the very
inner DM density distribution from our benchmark
$\alpha$-profile (solid line in Fig.~1, top panel). For
example, the process of galaxy formation could lead either to
flattening or to some steepening of the inner DM distribution.
The former may occur either owing to transfer of energy and/or
angular momentum from the baryons to the DM
\citep[see][]{El-Zant2001, Tonini2006}, or owing to quick mass
removal following the energy feedback from stars or active
galactic nuclei \citep[see discussion by][]{Lauer2007,
Kormendy2009}. On the other hand, steepening might be induced
by the `adiabatic' contraction of the baryons into the disc
\citep[see][]{Blumenthal1986, Mo1998}; but even in extreme
cases \citep[see discussion by][]{Abadi2009} such a contraction
would yield an inner DM density profile $\rho(r)\propto
r^{-3/(4-\gamma_a)}$, still \emph{flatter} than $1$ though
somewhat steeper than the original $\gamma_a\approx 0.63-0.75$.
Finally, at the very center of the Galaxy any accretion of DM
\citep[e.g.,][]{Gondolo1999, Bertone2002} onto the nuclear
supermassive black hole might enhance the DM distribution on
tiny scales $r< 10^{-1}$ pc.

Summing up, we stress that all such alterations of the inner
slope would occur on scales smaller than some $10$ pcs;
although significant at levels of a few percent to account for
the central stellar light\footnote{We note that a flat slope is
known to describe the very central light distribution in
luminous ellipticals, related to complex small-scale dynamics
\citep[see][]{Lauer2007, Kormendy2009}.}, their import is far
smaller for what the annihilation signal is concerned, and in
the average over $10^{-1}$ deg the flux is altered by less than
$0.1\%$. In fact, these corrections are currently at, or below
the resolution limit and the prospective sensitivity of
\textsl{Fermi}.

Other possible targets include the dwarf spheroidal galaxies in
the Local Group. These on the one hand constitute cleaner
environments than the GC owing to their dearth of stellar
sources; on the other hand, their distance if modest on
intergalactic scales, already makes detecting and resolving the
related annihilation signal a real challenge for \textsl{Fermi}
\citep[e.g.,][]{Pieri2009}. In addition, the shallow
gravitational potential wells of these systems make them
particularly prone to energy feedback events (see above), that
may flatten the inner DM distribution to flat slopes $\gamma_a<
0.63$ (consistent with kinematical observations), to the effect
of further lowering the annihilation signals. Upper limits more
stringent than the current value $\langle\Sigma v\rangle <
10^{-25}$ cm$^3$ s$^{-1}$ at a mass $m_{\rm DM}\approx 50$ GeV
will require delicate stacking over an ensemble of dwarfs.

Concerning particle cross section and masses, we recall that
the \textsl{PAMELA} satellite recently observed an excess of
the positron fraction $e^+/(e^+ + e^-)$ in the cosmic ray
$e^{\pm}$ spectra relative to the expected astrophysical
background above $10$ GeV \citep[see][]{Adriani2009}. This
excess can be simply explained in terms of a single or a few
sources like pulsars, that are expected to produce a powerlaw
spectrum of $e^{\pm}$ pairs with a cutoff at several TeVs
\citep[see][]{Bertone2009book}. On the other hand, the signal
may be also interpreted in terms of DM annihilations occurring
throughout the Galactic halo \citep[e.g.,][]{Bertone2009}.

If this is to be the case, however, the flux measured by
\textsl{PAMELA} mandates for very large effective annihilation
cross sections $\langle\Sigma v\rangle\sim 10^{-23}$ cm$^3$
s$^{-1}$, well above the natural value suggested by the
cosmological DM abundance (see \S~3). From a microphysical
point of view, this is still conceivable in scenarios with
Sommerfeld enhancements \citep[see discussion
by][]{Arkani-Hamed2009}; the cross section may be enhanced by a
factor $\sim 10^2$ for velocities $v/c\sim 10^{-3}$. On the
other hand, such a large Sommerfeld effects would also yield a
strong $\gamma$-ray annihilation signal towards the GC; for
this, little room is allowed on the basis of the current upper
limit provided by \textsl{Fermi} (see \S~3), unless the DM
particle mass substantially exceeds $50$ GeV.

Another possibility is to invoke a large boost factor of the
effective cross section due to clumpiness in the Galactic halo,
i.e., a crowd of dense subhalos; however, state-of-the-art
numerical simulations suggest such boosts not to be realistic
in the Galaxy, even less at the GC \citep[see][and discussion
by \citealt{Lattanzi2009}]{Springel2008}.

To sum up, we have discussed why the $\alpha$-profile with
$\alpha=1.25$ (see Fig.~1) constitutes a reliable DM
distribution in the Galaxy; we have argued that it will provide
a \emph{benchmark} to gauge in terms of DM annihilation the
$\gamma$ rays from the GC to be detected with \textsl{Fermi}
(see Fig.~2). Such an $\alpha$-profile will be instrumental to
derive reliable information concerning the microscopic nature
of the DM particles.

\begin{acknowledgements}
Work supported by Agenzia Spaziale Italiana (ASI), Istituto
Nazionale di Astrofisica (INAF) and Istituto Nazionale di
Fisica Nucleare (INFN). We acknowledge our referee S.H. Hansen
for keen and helpful comments. We are indebted to L. Ciotti, A.
Cirelli, P. Salucci, and M. Tavani for useful discussions. We
thank F. Vagnetti for critical reading. A. Lapi thanks
SISSA/ISAS and INAF-OATS for warm hospitality.
\end{acknowledgements}

\begin{appendix}

\section{Analytic fit to the $\alpha$-profile}

To complement the analytical details extensively dealt with by
\citet{Lapi2009a} and to enable a straightforward comparison
with the classic NFW and Einasto density runs, here we provide
a handy analytic fit to the $\alpha$-profiles in terms of the
deprojected S\'{e}rsic formula substantiated with parameters
directly \emph{derived} from the Jeans equation. We base on the
expression \citep[see][]{Prugniel1997}
\begin{equation}
{\rho(r)\over \rho(r_{-2})}=\left({r\over
r_{-2}}\right)^{-\tau}\,\exp{\left\{-{2-\tau\over
\eta}\,\left[\left({r\over r_{-2}}\right)^\eta-1\right]\right\}}~,
\end{equation}
where $\tau$ and $\eta$ are two fitting parameters; the
standard Einasto profile obtains for $\tau=0$. The values of
$\tau$ and $\eta$ for different $\alpha$ of interest here are
reported in Tables~A.1 and A.2 both in the isotropic and the
anisotropic cases; note that $\tau\approx\gamma_a$ is required
by the physical boundary condition satisfied at the center (see
\S~2). The resulting fits to the density runs of the
$\alpha$-profiles hold to better than $10\%$ in the relevant
range $10^{-2}\,r_{-2}\lesssim r\lesssim 10\,r_{-2}$.

The mass corresponding to the density distribution of Eq.~(A1)
reads
\begin{equation}
{M(<r)\over M_{\infty}}=\Gamma \left[{3-\tau\over\eta};
{2-\tau\over\eta}\,\left({r\over r_{-2}}\right)^{\eta}\right]~,
\end{equation}
where $\Gamma[a,x]\equiv \int_0^x{\rm
d}t\,t^{a-1}\,e^{-t}\big/\int_0^\infty{\rm
d}t\,t^{a-1}\,e^{-t}$ is the (normalized) incomplete
$\Gamma$-function.

\begin{table}
\caption{Values of the fitting parameters of Eq.~(A1) in the
isotropic case; $\alpha=1.25$ applies for the Galactic halo.}
\centering
\begin{tabular}{cccccc}
\hline\hline
$\alpha$ & $1.25$ & $1.26$ & $1.27$ \\
\hline
\\
$\tau$ & $0.750$ & $0.756$ & $0.762$ \\
\\
$\eta$ & $0.319$ & $0.298$ & $0.277$ \\
\\
\hline
\end{tabular}
\end{table}

\begin{table}
\caption{Values of the fitting parameters of Eq.~(A1) in the
anisotropic case; $\alpha=1.25$ applies for the Galactic halo.}
\centering
\begin{tabular}{cccccc}
\hline\hline
$\alpha$ & $1.25$ & $1.26$ & $1.27$ \\
\hline
\\
$\tau$ & $0.630$ & $0.636$ & $0.642$ \\
\\
$\eta$ & $0.364$ & $0.342$ & $0.319$ \\
\\
\hline
\end{tabular}
\end{table}

\end{appendix}


\begin{thebibliography}{63}

\bibitem[Abadi et al.(2009)]{Abadi2009}Abadi, M.G., Navarro,
    F.N., Fardal, M., Babul, A., and Steinmetz, M. 2009, MNRAS, submitted
    (preprint arXiv:0902.2477)

\bibitem[Abdo et al.(2009)]{Abdo2009}Abdo, A.A., et al. 2009,
    ApJS, 183, 46

\bibitem[Adriani et al.(2009)]{Adriani2009}Adriani, O., et al.
    [PAMELA Collaboration] 2009, Nature, 458, 607

\bibitem[Arkani-Hamed et
    al.(2009)]{Arkani-Hamed2009}Arkani-Hamed,
    N., Finkbeiner, D.P., Slatyer, T., and Weiner, N. 2009, Phys. Rev. D, 79, 015014

\bibitem[Ascasibar \& Gottl\"{o}ber(2008)]
    {Ascasibar2008}Ascasibar, Y., and Gottl\"{o}ber, S. 2008,
    ApJ, 386, 2022

\bibitem[Atwood et al.(2009)]{Atwood2009}Atwood, W.B., et al.
    2009, ApJ, 697, 1071

\bibitem[Austin et al.(2005)]{Austin2005}Austin, C.G., et al.
    2005, ApJ, 634, 756

\bibitem[Baer \& Tata(2009)]{Baer2009}Baer, H., and Tata, X.
    2009, in LHC physics, eds. A. Datta, B. Mukhopadhyaya and A. Raychaudhuri, in
    press (preprint arXiv:0805.1905)

\bibitem[Baltz et al.(2008)]{Baltz2008}Baltz, E.A., et al.
    2008,
    JCAP, 7, 13

\bibitem[Barbieri et al.(1982)]{Barbieri1982}Barbieri, R.,
    Ferrara, S., and Savoy, C. 1982, Phys. Lett. B, 119, 343

\bibitem[Bergstr\"{o}m(2009)]{Bergstrom2009}Bergstr\"{o}m, L.
    2009, in Dark Matter and Particle Physics (preprint arXiv:0903.4849)

\bibitem[Bergstr\"{o}m et al.
    (1998)]{Bergstrom1998}Bergstr\"{o}m,
    L., Ullio, P., and Buckley, J.H. 1998, Astropart. Phys., 9, 137

\bibitem[Bernabei et al.(2008)]{Bernabei2009}Bernabei,
    R., et al. 2008, Europ. Phys. J. C, 56, 333

\bibitem[Bertone(2009)]{Bertone2009book}Bertone, G.
    (\textit{ed.}), Particle Dark Matter (2009, Canbridge: Cambridge Univ. Press)

\bibitem[Bertone et al.(2009)]{Bertone2009}Bertone, G.,
    Cirelli,
    M.,  Strumia, A., and Taoso, M. 2009, JCAP, 3, 9

\bibitem[Bertone et al.(2005)]{Bertone2005}Bertone, G., Hooper,
    D.,  and Silk, J. 2005, Phys. Rept., 405, 279

\bibitem[Bertone et al.(2002)]{Bertone2002}Bertone, G., Sigl,
    G., and Silk, J. 2002, MNRAS, 337, 98

\bibitem[Binney(1978)]{Binney1978}Binney J. 1978, MNRAS, 183,
    779

\bibitem[Blumenthal et al.(1986)]{Blumenthal1986}Blumenthal,
    G.R., Faber, S.M., Flores, R., and Primack, J.R. 1986, ApJ, 301, 27

\bibitem[Bullock et al.(2001)]{Bullock2001}Bullock, J.S., et
    al.
    2001, MNRAS, 321, 559

\bibitem[Cavaliere et al.(2009)]{Cavaliere2009}Cavaliere, A.,
    Lapi, A.,  and Fusco-Femiano, R. 2009, ApJ, 698, 580

\bibitem[Cesarini et al.(2004)]{Cesarini2004}Cesarini, A.,
    Fucito, F., Lionetto, A., Morselli, A., and Ullio, P. 2004,
    Astrop. Phys., 21, 3, p. 267

\bibitem[Chamseddine et al.(1982)]{Chamseddine1982}Chamseddine,
    A., Arnowitt, R., and Nath, P. 1982, Phys. Rev. Lett., 49, 970

\bibitem[Dehnen \& McLaughlin(2005)]{Dehnen2005}Dehnen, W., and
    McLaughlin, D.E. 2005, MNRAS, 363, 1057

\bibitem[Diemand et al.(2007)]{Diemand2007}Diemand, J., Kuhlen,
    M., and Madau, P. 2007, ApJ, 667, 859

\bibitem[Diemand et al.(2005)]{Diemand2005}Diemand, J., et al.
    2005, MNRAS, 364, 665

\bibitem[El-Zant et al.(2001)]{El-Zant2001}El-Zant, A.,
    Shlosman, I., and Hoffman, Y. 2001, ApJ, 560, 636

\bibitem[Fornengo et al.(2004)]{Fornengo2004}Fornengo, N.,
    Pieri, L., and Scopel, S. 2004, Phys. Rev. D, 70, 3529.

\bibitem[Gondolo et al.(1999)]{Gondolo1999}Gondolo, P., and
    Silk, J. 1999, Phys. Rev. Lett. 83, 1719

\bibitem[Goodenough \&
    Hooper(2009)]{Goodenough2009}Goodenough,
    L., and Hooper, D. 2009, Fermilab-pub-09-494-A (preprint arXiv:0910.2998)

\bibitem[Graham et al.(2006)]{Graham2006}Graham, A.W., et al.
    2006, Astron. J., 132, 2685

\bibitem[Hall et al.(1983)]{Hall1983}Hall, L., Lykken, J., and
    Weinberg, S. 1983, Phys. Rev. D, 27, 2359

\bibitem[Hansen \& Moore(2006)]{Hansen2006}Hansen, S.H., and
    Moore, B. 2006, NewA, 11, 333

\bibitem[Hansen(2004)]{Hansen2004}Hansen, S.H.
    2004, MNRAS, 352, L41

\bibitem[Hansen(2007)]{Hansen2007}Hansen, B.V. 2007,
    B.S. thesis, Univ. Copenhagen

\bibitem[Hoffman et al.(2007)]{Hoffman2007}Hoffman, Y.,
    Romano-D\'iaz, E., Shlosman, I., and Heller, C. 2007, ApJ, 671, 1108

\bibitem[Kormendy et al.(2009)]{Kormendy2009}Kormendy, J., et
    al. 2009, ApJS, 182, 216

\bibitem[Lapi \& Cavaliere(2009a)]{Lapi2009a}Lapi, A., and
    Cavaliere, A. 2009a, ApJ, 692, 174

\bibitem[Lapi \& Cavaliere(2009b)]{Lapi2009b}Lapi, A., and
    Cavaliere, A. 2009b, ApJ, 695, L125

\bibitem[Lapi \& Cavaliere(2009c)]{Lapi2009c}Lapi, A.,
    and Cavaliere, A. 2009c, MNRAS, submitted

\bibitem[Lattanzi \& Silk(2009)]{Lattanzi2009}Lattanzi, M., and
    Silk, J. 2009, Phys. Rev. D, 79, h3523

\bibitem[Lauer et al.(2007)]{Lauer2007}Lauer, T.R., et al.
    2007, ApJ, 664, 226

\bibitem[Li et al.(2007)]{Li2007}Li, Y., Mo, H.J., van den
    Bosch, F.C., and Lin, W.P. 2007, MNRAS, 379, 689

\bibitem[Mo et al.(1998)]{Mo1998}Mo, H.J., Mao, S., and White,
    S.D.M. 1998, MNRAS, 295, 319

\bibitem[Navarro et al.(2008)]{Navarro2008}Navarro,
    J.F., et al. 2008, MNRAS, submitted (preprint arXiv:0810.1522)

\bibitem[Navarro et al.(1997)]{Navarro1997}Navarro, J.F.,
    Frenk, C.S., and White, S.D.M. 1997, ApJ, 490, 493

\bibitem[Ohta et al.(1983)]{Ohta1983}Ohta, N. 1983, Prog.
    Theor.
    Phys., 70, 542

\bibitem[Papucci \& Strumia(2009)]{Papucci2009}Papucci, M., and
    Strumia, A. 2009, CERN-PH-TH/2009-238 (preprint arXiv:0912.0742)

\bibitem[Peebles et al.(1983)]{Peebles1993}Peebles, P.J.E.
    1993,
    Principles of Physical Cosmology, (Princeton, NJ: Princeton Univ. Press)

\bibitem[Pieri et al.(2009)]{Pieri2009}Pieri, L., Pizzella, A.,
    Corsini, E.M., Dalla Bont\'a, E., and Bertola, F. 2009, A\&A, 496, 351

\bibitem[Prugniel \& Simien(1997)]{Prugniel1997}Prugniel, Ph.,
    and Simien, F. 1997, A\&A, 321, 111

\bibitem[Rasia et al.(2004)]{Rasia2004}Rasia, E., Tormen, G.,
    and Moscardini, L. 2004, MNRAS, 351, 237

\bibitem[Salvador-Sol\'e et al.(2007)]
    {Salvador2007}Salvador-Sol\'e, E., Manrique, A., Gonz\'alez-Casado,
    G., and Hansen, S.H. 2007, ApJ, 666, 181

\bibitem[Serpico \& Hooper(2009)]{Serpico2009}Serpico, P.D.,
    and Hooper, D. 2009, in Dark Matter and Particle Physics,
    in press (preprint arXiv:0902.2539)

\bibitem[Dunkley et al.(2009)]{Dunkley2009}Dunkley, J. et al.
    2009, ApJ, 180, 306

\bibitem[Schmidt et al.(2008)]{Schmidt2008}Schmidt,
    K.B., Hansen, S.H., \& Macci\'o, A.V. 2008, ApJ, 689, L33

\bibitem[Springel et al.(2008)]{Springel2008}Springel, V., et
    al. 2008, Nature, 456, 73

\bibitem[Springel et al.(2006)]{Springel2006}Springel, V.,
    Frenk, C.S., and White, S.D.M. 2006, Nature, 440, 1137

\bibitem[Striani(2009)]{Striani2009}Striani, E. 2009, B.S.
    Thesis, Univ. `Tor Vergata', Rome

\bibitem[Strigari(2007)]{Strigari2007}Strigari, L.E. 2007,
    Phys.
    Rev. D, 75, 3526

\bibitem[Taylor \& Navarro(2001)]{Taylor2001}Taylor, J.E., and
    Navarro, J.F. 2001, ApJ, 563, 483

\bibitem[Tonini et al.(2006)]{Tonini2006}Tonini, C., Lapi, A.,
    and Salucci, P. 2006, ApJ, 649, 591

\bibitem[Vass et al.(2008)]{Vass2008}Vass, I., Valluri, M.,
    Kravtsov, A., and Kazantzidis, S. 2008, MNRAS, 395, 1225

\bibitem[Vitale et al.(2009a)]{Vitale2009}Vitale, V., Morselli,
    A., et al. 2009, AIP Conf. Proc. 112, p. 164

\bibitem[Vitale et al.(2009b)]{Vitale2010}Vitale, V.,
    and Morselli, A. 2009, Poster presented at the
    Fermi Symposium, 2-5 November 2009, Washington DC (see
    \textsl{http://fermi.gsfc.nasa.gov/science/symposium/2009/})

\bibitem[Wechsler et al.(2006)]{Wechsler2006}Wechsler, R.H., et
    al. 2006, ApJ, 652, 71

\bibitem[White(1986)]{White1986}White, S.D.M. 1986, in Inner
    Space/Outer Space: the Interface between Cosmology and Particle Physics (Chicago:
    Chicago Univ. Press), p. 228-245

\bibitem[Zhao et al.(2003)]{Zhao2003}Zhao, D.H., Mo, H.J.,
    Jing,
    Y.P., and B\"{o}rner, G. 2003, MNRAS, 339, 12

\end{thebibliography}
\end{document}